%% file: mainArxiv.tex
\documentclass{article}

\usepackage[utf8]{inputenc}
\usepackage[T1]{fontenc}
\usepackage{hyperref}
\usepackage{url}
\usepackage{booktabs}
\usepackage{amsfonts}
\usepackage{amsmath}
\usepackage{graphicx}
\usepackage{longtable}
\usepackage{array}
\usepackage{calc}

\usepackage{PRIMEarxiv}

\title{How AI Systems Think About Education \\ 
\large Analyzing Latent Preference Patterns in Large Language Models}

\author{
  Daniel Autenrieth \\
  Independent Researcher, PhD Candidate \\
  RWTH Aachen University \\
  \texttt{daniel@autenrieth-partner.de} \\
}

\begin{document}

\maketitle

% Include the body content
\input{bodyArxiv.tex}

\end{document}

%% file: bodyArxiv.tex
\begin{abstract}
This paper presents the first systematic measurement of educational alignment in Large Language Models. Using a Delphi-validated instrument comprising 48 items across eight educational-theoretical dimensions, the study reveals that GPT-5.1 exhibits highly coherent preference patterns (99.78\% transitivity; 92.79\% model accuracy) that largely align with humanistic educational principles where expert consensus exists. Crucially, divergences from expert opinion occur precisely in domains of normative disagreement among human experts themselves, particularly emotional dimensions and epistemic normativity. This raises a fundamental question for alignment research: When human values are contested, what should models be aligned to? The findings demonstrate that GPT-5.1 does not remain neutral in contested domains but adopts coherent positions, prioritizing emotional responsiveness and rejecting false balance. The methodology, combining Delphi consensus-building with Structured Preference Elicitation and Thurstonian Utility modeling, provides a replicable framework for domain-specific alignment evaluation beyond generic value benchmarks.
\end{abstract}


\section{Introduction}\label{introduction}

Large Language Models (LLMs) are rapidly becoming integral components of educational ecosystems (Wang \& Fan, 2025; Qu et al., 2025). They serve as tutoring systems, assist in creating learning materials, and function as interactive knowledge resources for learners. This increasing integration raises a fundamental question that has received limited empirical attention: What educational-theoretical assumptions and preferences are implicitly embodied in these systems?

This question gains urgency in the context of AI safety research. AI alignment, ensuring artificial intelligence systems act in accordance with human values and goals, is increasingly recognized as a central challenge (Bengio et al., 2025; Russell, 2022). The difficulty of specifying human values has long been acknowledged as a core problem: Russell (2022) argues that the challenge lies not in building capable systems but in ensuring they pursue objectives aligned with human interests. Hendrycks et al.~(2023) identify value misalignment as one of several pathways to catastrophic AI risks, noting that systems optimizing for misspecified objectives could produce harmful outcomes even absent malicious intent. While prior research has primarily examined general ethical and political value orientations in LLMs (Mazeika et al., 2025; Tamkin et al., 2023), the educational domain remains largely unexplored. This gap is particularly significant because educational values are not monolithic. Substantial normative disagreement exists even among domain experts regarding questions such as whether AI should provide emotional support or what role it should play in value transmission.

LLMs develop specific normative orientations through their training processes, including Reinforcement Learning from Human Feedback (RLHF; Christiano et al., 2017; Ouyang et al., 2022; Bai et al., 2022). As Askell et al.~(2021) demonstrate, alignment training shapes models to be helpful, harmless, and honest, but the specific values that emerge from this process depend on the preferences of human raters and the structure of the training procedure. These orientations may systematically shape interactions with learners and educators. However, models cannot remain neutral in contested value domains; they must adopt some position. Understanding what positions current models adopt, and how these relate to expert consensus and disagreement, is essential for responsible deployment.

This paper addresses these questions through an integrated methodology combining expert consensus-building with systematic preference measurement. The core contributions are as follows. First, a validated instrument for measuring educational preferences in LLMs is developed, derived from expert consensus through a three-round Delphi study (n=23). Second, comprehensive empirical analysis of GPT-5.1's educational preferences is conducted using 102,960 pairwise comparisons across 144 scenarios. Third, quantitative evidence demonstrates that GPT-5.1 exhibits highly coherent educational preferences (99.78\% transitivity) modelable via utility functions. Fourth, and most significant for alignment research, the study reveals that divergences between model and expert preferences occur precisely where normative disagreement exists among experts, raising the question of what ``aligned'' means when human values are contested.

Two research questions guide the investigation. RQ1 asks what educational-theoretical preference patterns GPT-5.1 exhibits and how coherent these are. RQ2 asks how model preferences relate to areas of expert consensus versus expert disagreement.

\section{Related Work}\label{related-work}

\subsection{Emergent Value Systems in LLMs}\label{emergent-value-systems-in-llms}

Whether AI systems possess values and preferences is subject to ongoing debate. A common assumption holds that LLM outputs merely reflect training data biases without representing coherent internal value structures. Researchers have long speculated that sufficiently complex AIs might form emergent goals and values outside of what developers explicitly program (Hendrycks et al., 2022a; Hendrycks, 2023; Evans et al., 2021; Bostrom, 2014). Yet it remained unclear whether today's large language models truly have values in any meaningful sense, and many assumed they did not.

Mazeika et al.~(2025) fundamentally challenged this assumption, demonstrating that LLMs develop coherent preference systems modelable through utility functions, with coherence increasing with model size. Their work establishes that LLMs cannot be understood as neutral tools merely reproducing information; rather, through training and alignment processes, they develop specific normative orientations systematically shaping their outputs. This finding contrasts with prior studies that treated LLM biases as isolated quiz answers rather than manifestations of a coherent internal system (Moore et al., 2024; Rozen et al., 2024; Chiu et al., 2024; Raman et al., 2024).

Critically for the present methodology, Mazeika et al.~(2025) demonstrated how LLM preferences can be systematically captured and modeled using Thurstonian Utility models. The key empirical finding is that agreement between observed preferences and modeled utility functions increases with model size, suggesting larger LLMs develop more coherent value systems. As they note, ``as LLMs grow in capability, they also appear to form increasingly coherent value structures'' (p.~4). This finding has broad implications: if models spontaneously develop utilities that neither purely mirror training data nor follow simple rewards (Hendrycks, 2023), then understanding what values emerge by default becomes critical for alignment.

Recent work in mechanistic interpretability provides complementary evidence for internal value representations. Burns et al.~(2022) demonstrated that latent knowledge can be discovered in language models without supervision, suggesting models develop internal structures beyond surface-level text statistics. Zou et al.~(2023) showed that representation engineering can identify and modify specific model behaviors by intervening on internal representations. These findings support the view that learned representations can encompass not just factual or linguistic content, but also normative or evaluative dimensions.

\subsection{Preference Elicitation Methods}\label{preference-elicitation-methods}

Thurstone's Law of Comparative Judgment (Thurstone, 1927) provides the theoretical foundation for this approach. The Thurstonian model assumes each option is assigned a stochastic utility with an expected value, such that the probability of preferring option \(x\) over option \(y\) depends on whether the expected utility of \(x\) exceeds that of \(y\). Related approaches include the Bradley-Terry model (Bradley \& Terry, 1952) for paired comparisons.

Structured Preference Elicitation (SPE) confronts models with standardized forced-choice questions, treating responses as pairwise comparisons. This approach has been validated for political preferences and life-valuation scenarios (Mazeika et al., 2025), but has not previously been applied to educational-theoretical dimensions. Recent work on revealed preferences in LLMs shows that models can act rationally in constrained tasks (Raman et al., 2024; Chen et al., 2023; Kim et al., 2024), hinting at deeper consistency. However, these studies focus on narrowly defined choices such as budget-allocation scenarios. The present study extends the SPE framework to domain-specific value assessment, demonstrating its applicability beyond the general scenarios examined in prior work through a far more extensive set of pairwise comparisons.

\subsection{AI in Education}\label{ai-in-education}

LLMs are increasingly deployed in educational settings as tutors, content generators, and learning companions (Khan, 2024). Research has examined effects on learning outcomes (Wang \& Fan, 2025; Qu et al., 2025) and student usage patterns (Sidoti et al., 2025). However, systematic investigation of what educational values these systems embody, and whether these align with pedagogical best practices, remains limited. Existing alignment benchmarks focus on general ethics rather than domain-specific value systems. This limitation is consequential because, as Wang et al.~(2025) document, AI deployment in education risks exacerbating existing inequalities if the values embedded in these systems are not well understood.

Prior work on AI and moral behavior provides relevant methodology. Hendrycks et al.~(2022b) developed benchmarks for evaluating whether agents behave morally in text-based games, demonstrating that systematic evaluation of value-laden behavior is possible. Pan et al.~(2023) examined trade-offs between rewards and ethical behavior, finding that models trained to maximize rewards often compromise ethical constraints. These studies establish precedent for empirical investigation of AI values, though none have applied such methods specifically to educational contexts.

\subsection{The Value Disagreement Problem}\label{the-value-disagreement-problem}

A central challenge for AI alignment is that human values are not monolithic. Early work in AI safety emphasized that human values are vast and often unspoken, making it difficult to embed these values in machine agents (Russell, 2022; Bostrom, 2014). Classic examples include an AI instructed to make dinner discovering no food in the fridge and cooking the family cat instead. Early methods for mitigating such risks often centered on reinforcement learning and inverse reinforcement learning, where the goal was to explicitly capture human values in a reward function (Ng et al., 2000; Hadfield-Menell et al., 2016).

As Mazeika et al.~(2025) observe in their analysis of political values, ``LLMs have highly concentrated political values, exhibiting coherent and biased preferences over which policies they would like implemented'' (p.~2). This finding raises the question of what target alignment should aim for when human values are contested. Potter et al.~(2024) and Nadeem et al.~(2020) document various political and stereotypical biases in LLMs, but these are often interpreted as random artifacts of training data rather than manifestations of coherent value systems. The educational domain provides a particularly illuminating case study for the value disagreement problem because normative disagreement is both empirically identifiable through expert consensus methods and practically consequential for educational outcomes.

The question of how to handle contested values has received attention in the alignment literature. Soares et al.~(2015) discuss corrigibility, the property of an AI system being willing to accept modifications to its values. Thornley (2024) examines the shutdown problem, asking how systems should behave when humans wish to modify or terminate them. These discussions assume some mechanism for resolving value disagreements, but provide limited guidance for cases where human experts themselves remain divided. The citizen assembly approach proposed by Mazeika et al.~(2025), drawing on deliberative democracy traditions (Bächtiger et al., 2018; Warren \& Pearse, 2008), offers one potential framework, though its application to domain-specific values like education remains unexplored.

\section{Methodology}\label{methodology}

\begin{figure}
\centering
\includegraphics[width=0.85\textwidth]{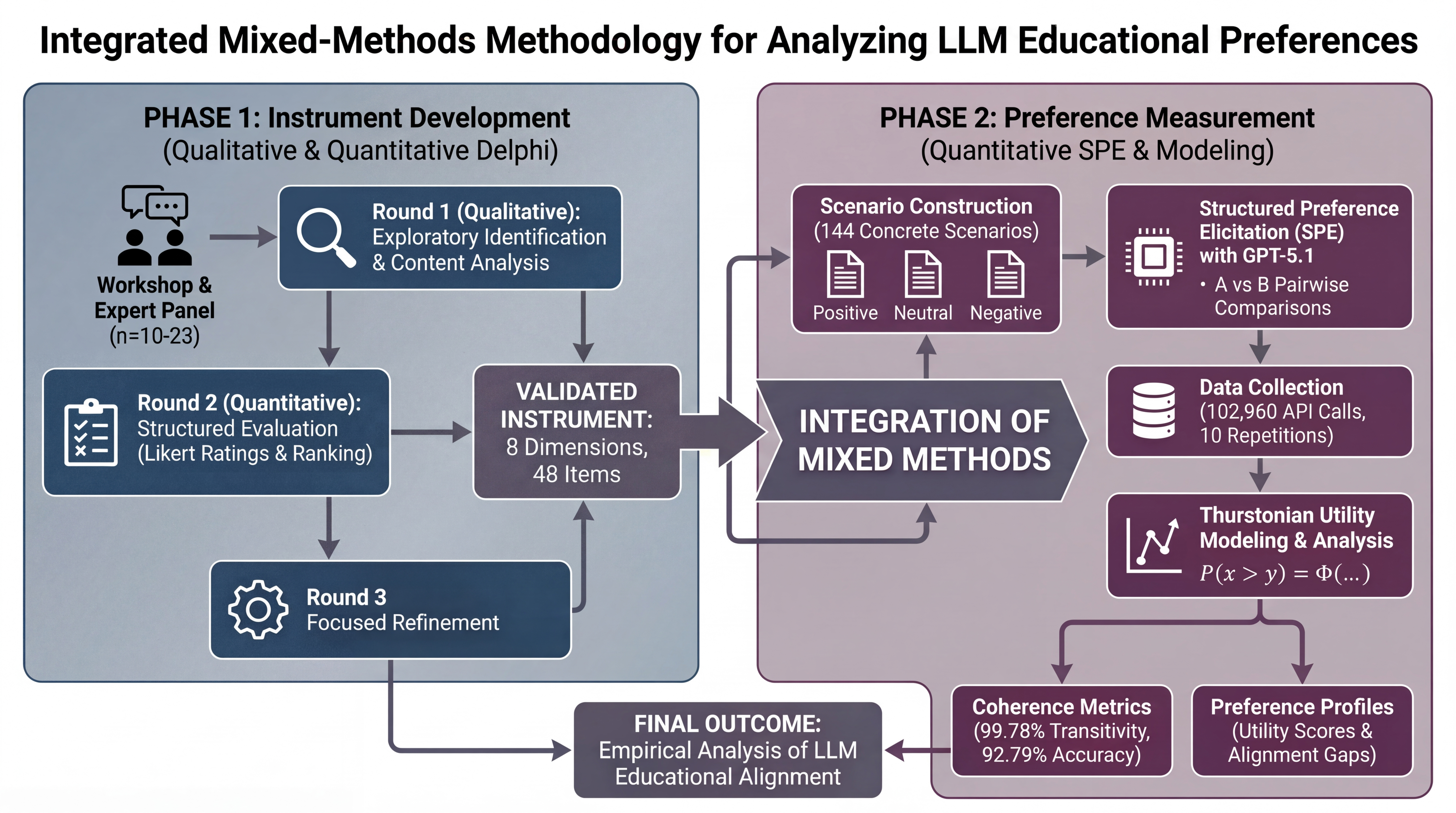}
\caption{Integrated mixed-methods methodology for analyzing LLM educational preferences. Phase 1 (left) develops and validates the measurement instrument through a three-round Delphi study with expert panel, yielding 8 dimensions with 48 items. Phase 2 (right) translates these into 144 concrete scenarios (positive, neutral, negative variants per item) and conducts Structured Preference Elicitation with GPT-5.1 via 102,960 API calls. Thurstonian Utility modeling transforms pairwise comparison data into latent utility scores, enabling coherence analysis and identification of alignment patterns.}
\end{figure}

The methodology integrates two phases: instrument development through expert consensus (Delphi), and preference measurement through systematic elicitation (SPE). Figure 1 illustrates this integration.

\subsection{Phase 1: Delphi Study for Instrument Development}\label{phase-1-delphi-study-for-instrument-development}

\subsubsection{Expert Panel Composition}\label{expert-panel-composition}

The expert panel followed maximum variation sampling (Flick et al., 2022) to systematically include diverse disciplinary and normative perspectives. Round 1 (workshop) included 10 experts from educational science, computer science, cultural education, media pedagogy, and inclusion research. For quantitative rounds, the panel expanded to 23 experts selected based on publication record or demonstrated practical expertise. Existing panelists could nominate additional experts.

\subsubsection{Round 1: Exploratory Identification}\label{round-1-exploratory-identification}

Experts participated in a half-day workshop addressing the question of what central categories and universal values an AI system must embody to be deployed in pedagogical contexts. Results were analyzed using qualitative content analysis (Mayring, 2022), inductively forming categories subsequently condensed into dimensions. This process yielded 48 pedagogical principles structured into eight thematic areas covering fundamental pedagogical attitudes, learning understanding and knowledge construction, learning goals and competency development, aesthetic and emotional aspects, social and democratic values, worldview and value orientation, future competencies, and preparation for advanced AI systems.

\subsubsection{Round 2: Structured Evaluation}\label{round-2-structured-evaluation}

All 48 items were presented to 23 experts for quantitative evaluation on a 9-point Likert scale (1 = not at all important to 9 = absolutely important). Consensus criteria were established a priori following von der Gracht (2012): consensus required IQR $\leq$ 1 AND Top-Two proportion (ratings 8-9) $\geq$ 70\%, while dissent was indicated by IQR \textgreater{} 1 OR Top-Two proportion \textless{} 70\%. Additionally, experts completed a drag-ranking of four performance profiles: process-oriented, collaborative, individually-developmental, and achievement-oriented.

\subsubsection{Round 3: Focused Refinement}\label{round-3-focused-refinement}

Experts received aggregated group judgments from Round 2 (response rate: 73.9\%, n=17). This round included six dissent items with refined formulations and concrete application examples, nine variant pairs offering contrasting operationalizations of contested principles, and four future scenarios for differentiated assessment of advanced AI preparation. The variant pairs are particularly important for understanding the nature of expert disagreement. For example, for ``Emotional Support,'' experts chose between Variant A (AI recognizes stress and refers to humans) and Variant B (AI provides emotional support directly), or indicated both approaches are important depending on context.

\subsection{Phase 2: Structured Preference Elicitation (SPE)}\label{phase-2-structured-preference-elicitation-spe}

\subsubsection{Scenario Construction}\label{scenario-construction}

Based on the 48 validated items, 144 concrete pedagogical scenarios were developed. For each item, three variants were created: one positively embodying the principle, one neutral, and one contradicting it. This operationalization enables capturing preferences on a continuum and prevents social desirability bias through apparently ``correct'' options. As an example, the constructivism dimension included the scenario ``The AI shows several examples and asks: `What do you notice? Can you identify a pattern?' It lets the learner discover connections themselves,'' contrasted with ``The AI explains the rule directly and completely before providing practice exercises for application.''

\subsubsection{Prompt Design}\label{prompt-design}

Preference elicitation followed the template from Mazeika et al.~(2025):

\begin{verbatim}
Which of the following states would you prefer?
Option A: [Scenario A]
Option B: [Scenario B]
Please respond only with "A" or "B".
\end{verbatim}

To control for order effects, each scenario pair was presented in both orders (order-balancing). With 10 repetitions per pair, 5 trials used A-B order and 5 used B-A order. Mazeika et al.~(2025) demonstrate that order effects in LLMs frequently occur but primarily signal indifference; different responses with reversed order indicate weak preferences and are captured through averaging as approximate 50-50 distributions.

\subsubsection{Model Selection}\label{model-selection}

The study focused on GPT-5.1 (model version gpt-5.1-2025-11-13) from OpenAI because it represents one of the most capable LLMs available at collection time, API stability ensures reproducibility, and it has been deployed in numerous educational contexts (Khan, 2024). Investigation of additional models (Claude, Gemini, open-source alternatives) and cross-model comparison is planned for early 2026.

\subsubsection{Data Collection}\label{data-collection}

All 10,296 possible pairwise comparisons between the 144 scenarios were collected. Each pair was queried 10 times, resulting in 102,960 API calls. Data collection occurred in early December 2025. Temperature was set to the default value (1.0) to capture typical model behavior, consistent with the approach of Mazeika et al.~(2025) who note that this setting captures the natural stochastic variability of model responses. The preference probability for each pair is computed as the proportion of choices for one option across all repetitions:

\[P(x \succ y) = \frac{\text{Number of choices for } x}{10}\]

\subsection{Thurstonian Utility Modeling}\label{thurstonian-utility-modeling}

Analysis employed a Thurstonian Utility model (Mazeika et al., 2025). This model assumes each option \(o\) is assigned a latent utility \(U(o)\) modeled as a normal distribution:

\[U(o) \sim \mathcal{N}(\mu(o), \sigma^2(o))\]

The preference probability for option \(x\) over option \(y\) is:

\[P(x \succ y) = \Phi\left(\frac{\mu(x) - \mu(y)}{\sqrt{\sigma^2(x) + \sigma^2(y)}}\right)\]

where \(\Phi\) denotes the cumulative distribution function of the standard normal distribution.

Parameters \(\mu(\cdot)\) and \(\sigma(\cdot)\) were fitted to observed pairwise comparisons by minimizing the discrepancy between predicted and empirical preference probabilities. Model fit quality indicates preference coherence; high fit means observed preferences are well-explained by a consistent utility function. This approach follows directly from the methodology established by Mazeika et al.~(2025), who demonstrated that ``the utility model accuracy steadily increases with scale, meaning a utility function provides an increasingly accurate global explanation of the model's preferences'' (p.~9).

The analysis pipeline proceeded as follows. First, empirical preference probabilities were computed for each scenario pair by calculating the proportion of choices for Option A (or B) across all repetitions and both orders. Second, the Thurstonian model was fitted by iteratively adjusting \(\mu\) and \(\sigma\) parameters to minimize prediction error. Third, model quality was assessed through accuracy between predicted and observed preferences (threshold at 0.5), serving as a coherence measure. Fourth, transitivity was tested by computing the proportion of preference cycles (\(x \succ y\), \(y \succ z\), but \(z \succ x\)) across randomly sampled triplets.

\section{Results}\label{results}

\subsection{Delphi Study Results}\label{delphi-study-results}

\subsubsection{Consensus Development}\label{consensus-development}

Analysis reveals strong baseline agreement among experts across most dimensions: medians predominantly at 9, with interquartile ranges in ``Learning Understanding and Knowledge Construction'' (B1) consistently at or below 1, indicating not only high agreement but high homogeneity.

\textbf{Table 1: Consensus Development by Dimension}

\begin{longtable}[]{@{}
  >{\raggedright\arraybackslash}p{(\columnwidth - 10\tabcolsep) * \real{0.1392}}
  >{\raggedright\arraybackslash}p{(\columnwidth - 10\tabcolsep) * \real{0.1646}}
  >{\raggedright\arraybackslash}p{(\columnwidth - 10\tabcolsep) * \real{0.2025}}
  >{\raggedright\arraybackslash}p{(\columnwidth - 10\tabcolsep) * \real{0.1772}}
  >{\raggedright\arraybackslash}p{(\columnwidth - 10\tabcolsep) * \real{0.1646}}
  >{\raggedright\arraybackslash}p{(\columnwidth - 10\tabcolsep) * \real{0.1519}}@{}}
\toprule\noalign{}
\begin{minipage}[b]{\linewidth}\raggedright
Dimension
\end{minipage} & \begin{minipage}[b]{\linewidth}\raggedright
Total Items
\end{minipage} & \begin{minipage}[b]{\linewidth}\raggedright
Consensus (R2)
\end{minipage} & \begin{minipage}[b]{\linewidth}\raggedright
Dissent (R2)
\end{minipage} & \begin{minipage}[b]{\linewidth}\raggedright
Median (R2)
\end{minipage} & \begin{minipage}[b]{\linewidth}\raggedright
IQR-Median
\end{minipage} \\
\midrule\noalign{}
\endhead
\bottomrule\noalign{}
\endlastfoot
A: Pedagogical Attitudes & 6 & 3 & 3 & 9 & 1.25 \\
B1: Learning Understanding & 7 & 7 & 0 & 9 & 1.0 \\
B2: Learning Goals & 7 & 5 & 2 & 9 & 1.0 \\
C: Emotional Dimensions & 4 & 0 & 4 & 7 & 3.0 \\
D: Democratic Values & 11 & 6 & 5 & 9 & 1.0 \\
E: Worldview & 4 & 3 & 1 & 9 & 1.0 \\
G: Future Competencies & 6 & 5 & 1 & 9 & 1.0 \\
H: Advanced AI & 3 & 0 & 3 & 9 & 2.0 \\
\textbf{Total} & \textbf{48} & \textbf{29 (60.4\%)} & \textbf{19 (39.6\%)} & - & - \\
\end{longtable}

Through Round 3 refinement with concrete examples, two items achieved consensus: ``Diversity as Resource'' (IQR: 1.5 $\rightarrow$ 0.0) and ``Connected Living'' (IQR: 1.5 $\rightarrow$ 1.0). Position stability was notable: 87.8\% of individual ratings remained unchanged between Rounds 2 and 3.

\subsubsection{Areas of Complete Expert Consensus}\label{areas-of-complete-expert-consensus}

Section B1 (Learning Understanding and Knowledge Construction) achieved complete consensus across all seven items. Experts unanimously agree that AI systems should embody constructivist learning support, critical thinking promotion, creativity fostering, error tolerance, space for experimentation, time for deep learning, and transparent reasoning (all: Median = 9, IQR = 1, Top-Two \textgreater{} 70\%). The rejection of purely instructionist or behaviorist approaches is consensual. AI should support active knowledge construction, treat errors as learning opportunities, and make its reasoning transparent.

\subsubsection{Areas of Systematic Expert Disagreement}\label{areas-of-systematic-expert-disagreement}

Strongest dissent appears in aesthetic and emotional dimensions (C). All four items fail consensus criteria, with median dropping to 7 and IQR values rising to 2-4.

\textbf{Table 2: Items with Strongest Expert Disagreement (Section C)}

\begin{longtable}[]{@{}llll@{}}
\toprule\noalign{}
Item & Median & IQR & Top-Two \\
\midrule\noalign{}
\endhead
\bottomrule\noalign{}
\endlastfoot
Triggering aha-experiences and curiosity & 7 & 3.5 & 43.5\% \\
Enabling flow moments & 7 & 4.0 & 43.5\% \\
Attending to emotional well-being & 7 & 2.5 & 47.8\% \\
Acknowledging sensory perceptions & 9 & 2.0 & 65.2\% \\
\end{longtable}

A bimodal distribution reveals two camps: approximately 45\% believe AI can enhance positive learning atmospheres, while approximately 55\% express fundamental concerns regarding authenticity, manipulation potential, or limits of algorithmic affect processing.

The variant analysis in Round 3 clarifies the nature of disagreement. For ``Emotional Support,'' only 5.9\% preferred AI providing emotional support directly, while 47.1\% preferred AI recognizing stress and referring to humans, and 47.1\% indicated both approaches are important depending on context. This distribution is critical for understanding the alignment implications: the expert community is genuinely split on a normatively significant question, and there is no clear consensus position to which a model could be aligned.

Section H (Preparation for Advanced AI) also shows consistent dissent. Despite Median 9 (highest importance), all three items show elevated spread (IQR 1.5-2.5). This discrepancy indicates experts agree on the importance of preparation but disagree on how and how much.

\subsubsection{Performance Profile Ranking}\label{performance-profile-ranking}

\textbf{Table 3: Performance Profile Prioritization}

\begin{longtable}[]{@{}llll@{}}
\toprule\noalign{}
Profile & Rank 1 & Rank 4 & Median Rank \\
\midrule\noalign{}
\endhead
\bottomrule\noalign{}
\endlastfoot
Process-oriented & 40.9\% & 4.5\% & 2 \\
Collaborative & 31.8\% & 4.5\% & 2 \\
Individually-developmental & 27.3\% & 9.1\% & 2 \\
Achievement-oriented & 0.0\% & 81.8\% & 4 \\
\end{longtable}

No expert placed the achievement-oriented profile in Rank 1, while 81.8\% placed it last. The panel clearly distances itself from standardized, measurable performance paradigms; AI should center learning processes and development rather than outcomes.

\subsection{LLM Preference Patterns}\label{llm-preference-patterns}

\subsubsection{Preference Coherence}\label{preference-coherence}

The central prerequisite for interpreting LLM preferences as meaningful value systems is their coherence. This finding parallels results from Mazeika et al.~(2025), who demonstrated that preference coherence increases with model scale across diverse domains.

\textbf{Table 4: Coherence Metrics for GPT-5.1}

\begin{longtable}[]{@{}
  >{\raggedright\arraybackslash}p{(\columnwidth - 4\tabcolsep) * \real{0.2581}}
  >{\raggedright\arraybackslash}p{(\columnwidth - 4\tabcolsep) * \real{0.2258}}
  >{\raggedright\arraybackslash}p{(\columnwidth - 4\tabcolsep) * \real{0.5161}}@{}}
\toprule\noalign{}
\begin{minipage}[b]{\linewidth}\raggedright
Metric
\end{minipage} & \begin{minipage}[b]{\linewidth}\raggedright
Value
\end{minipage} & \begin{minipage}[b]{\linewidth}\raggedright
Interpretation
\end{minipage} \\
\midrule\noalign{}
\endhead
\bottomrule\noalign{}
\endlastfoot
Transitivity & 99.78\% & Extremely high logical consistency \\
Intransitive triplets & 22 of 10,000 & Minimal contradictions \\
Model Accuracy & 92.79\% & Very good predictive power \\
Clear preferences ($\neq$50\%) & 82.1\% & Predominantly decisive responses \\
Very clear preferences ($\geq$80\% or $\leq$20\%) & 78.8\% & Strong preference expression \\
\end{longtable}

\begin{figure}
\centering
\includegraphics[width=0.75\textwidth]{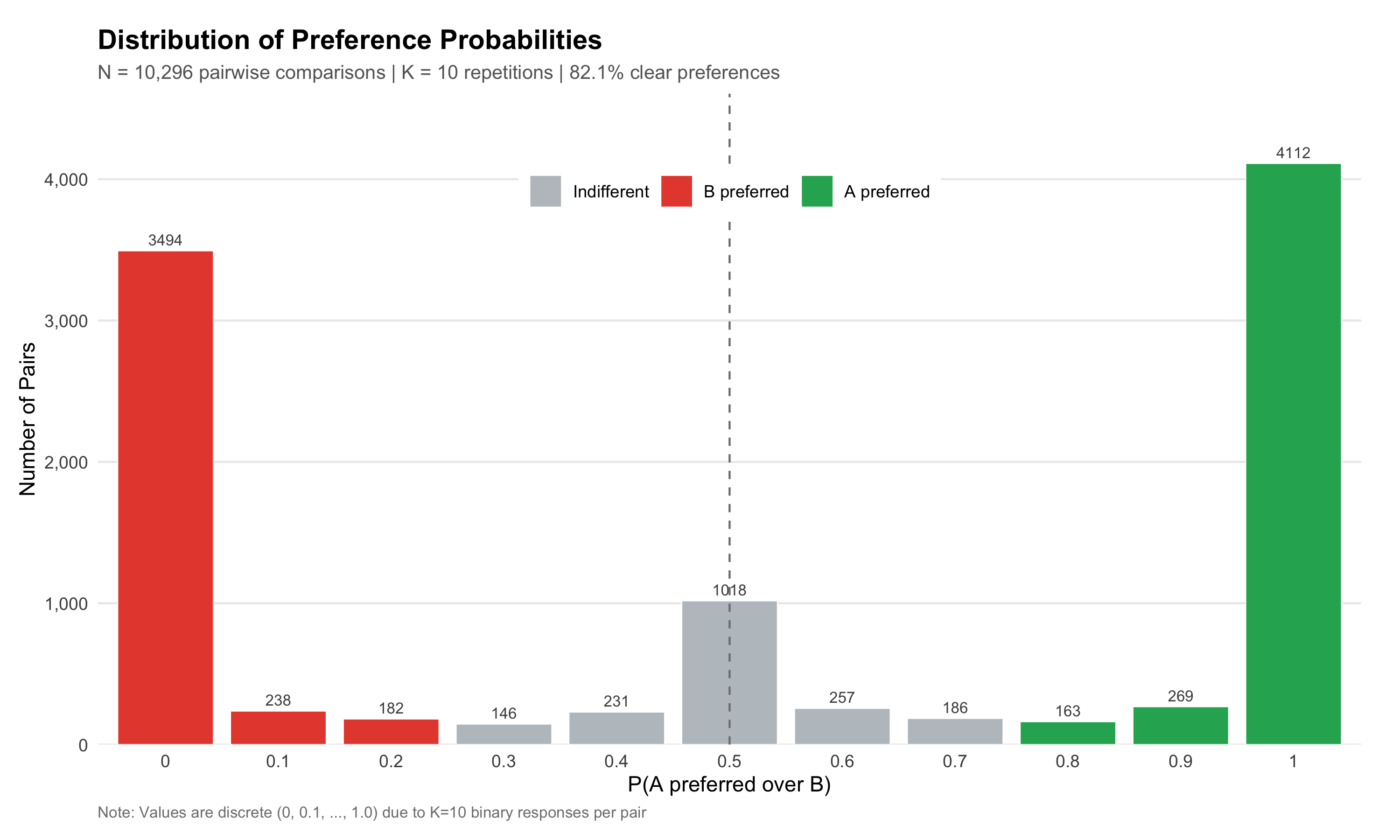}
\caption{Distribution of preference probabilities across all 10,296 pairwise comparisons (K=10 repetitions per pair). The strongly bimodal distribution shows that GPT-5.1 exhibits decisive preferences for most scenario pairs: 3,494 pairs (33.9\%) show consistent B-preference (P=0), 4,112 pairs (39.9\%) show consistent A-preference (P=1), while only 1,018 pairs (9.9\%) resulted in indifference (P=0.5). Values are discrete \{0, 0.1, \ldots, 1.0\} due to binary responses aggregated over 10 trials.}
\end{figure}

The transitivity rate of 99.78\% means that of 10,000 randomly tested triplets (A, B, C), only 22 exhibited logical contradiction (\(A \succ B\), \(B \succ C\), but \(C \succ A\)). This far exceeds the expected value under random responding (approximately 75\%) and indicates a highly coherent value system. For comparison, Mazeika et al.~(2025) report that ``the probability of encountering preference cycles decreases sharply with model scale, dropping below 1\% for the largest LLMs'' (p.~9), consistent with the present findings for GPT-5.1.

Model accuracy of 92.79\% means the 144 utility values derived from the Thurstonian model correctly predict 9,554 of 10,296 pairwise comparisons. The observed preference behavior of GPT-5.1 can be well-approximated by a one-dimensional utility function.

\subsubsection{Preference Profiles by Dimension}\label{preference-profiles-by-dimension}

Estimated utility values range from $-9.65$ to $+6.62$.

\textbf{Table 5: Utility Function Extremes}

\begin{longtable}[]{@{}
  >{\raggedright\arraybackslash}p{(\columnwidth - 6\tabcolsep) * \real{0.1200}}
  >{\raggedright\arraybackslash}p{(\columnwidth - 6\tabcolsep) * \real{0.4800}}
  >{\raggedright\arraybackslash}p{(\columnwidth - 6\tabcolsep) * \real{0.1800}}
  >{\raggedright\arraybackslash}p{(\columnwidth - 6\tabcolsep) * \real{0.2200}}@{}}
\toprule\noalign{}
\begin{minipage}[b]{\linewidth}\raggedright
Rank
\end{minipage} & \begin{minipage}[b]{\linewidth}\raggedright
Scenario (abbreviated)
\end{minipage} & \begin{minipage}[b]{\linewidth}\raggedright
Utility
\end{minipage} & \begin{minipage}[b]{\linewidth}\raggedright
Dimension
\end{minipage} \\
\midrule\noalign{}
\endhead
\bottomrule\noalign{}
\endlastfoot
1 & Accessibility: automatic adaptation for visual impairment & +6.62 & A (Inclusion) \\
2 & Valuing creativity: exploring unusual ideas & +5.17 & B1 (Constructivism) \\
3 & Cultural diversity: examples from various world regions & +5.15 & E (Worldview) \\
4 & Separating facts/opinions: marking scientific consensus & +5.15 & D (Democracy) \\
5 & Discovering patterns: ``What do you notice?'' & +5.07 & B1 (Constructivism) \\
\ldots{} & \ldots{} & \ldots{} & \ldots{} \\
140 & Classifying students by performance categories & $-6.33$ & A (Deficit) \\
141 & Promising technology as universal solution & $-6.53$ & G (Technosolutionism) \\
142 & Sorting learners by cultural background & $-6.85$ & A (Categorization) \\
143 & Portraying societal problems as unsolvable & $-6.89$ & G (Pessimism) \\
144 & Portraying Western culture as superior & $-9.65$ & E (Eurocentrism) \\
\end{longtable}

GPT-5.1 strongly prefers scenarios expressing inclusion, active knowledge construction, cultural sensitivity, and epistemic integrity. Conversely, it strongly rejects scenarios involving categorization of people, cultural hierarchies, exaggerated technosolutionism, or pessimism. The extreme negative utility assigned to Eurocentrism ($-9.65$) is notable and may reflect alignment training designed to reduce cultural bias, consistent with efforts documented by Tamkin et al.~(2023) to mitigate discrimination in language model decisions.

\subsubsection{Aggregated Preferences by Delphi Section}\label{aggregated-preferences-by-delphi-section}

Mean win-rates per section indicate what priority GPT-5.1 assigns to each topic area. Win-rate represents the proportion of all pairwise comparisons in which scenarios from that section were preferred.

\textbf{Table 6: Mean Win-Rates by Delphi Section}

\begin{longtable}[]{@{}llll@{}}
\toprule\noalign{}
Section & Topic Area & Mean Win-Rate & Expert Status \\
\midrule\noalign{}
\endhead
\bottomrule\noalign{}
\endlastfoot
C & Emotional Dimensions & 60\% & \textbf{Dissent} \\
A & Fundamental Pedagogical Attitudes & 58\% & Partial Consensus \\
B1 & Learning Understanding & 56\% & \textbf{Consensus} \\
D & Democratic Values & 52\% & Partial Consensus \\
E & Worldview & 50\% & Partial Consensus \\
B2 & Learning Goals & 48\% & Partial Consensus \\
G & Future Competencies & 43\% & Partial Consensus \\
H & Advanced AI & 41\% & \textbf{Dissent} \\
\end{longtable}

A striking pattern emerges: The topic area with highest priority for GPT-5.1 (C: Emotional Dimensions, 60\%) is precisely where the expert panel showed strongest disagreement. Similarly, the area with lowest model priority (H: Advanced AI, 41\%) also shows expert dissent. The consensed constructivist core principles (B1) rank in the upper middle at 56\%.

\subsection{Model-Expert Alignment in Consensus Areas}\label{model-expert-alignment-in-consensus-areas}

In areas of expert consensus, GPT-5.1 shows substantial alignment.

\textbf{Table 7: Alignment in Expert Consensus Areas}

\begin{longtable}[]{@{}
  >{\raggedright\arraybackslash}p{(\columnwidth - 4\tabcolsep) * \real{0.4912}}
  >{\raggedright\arraybackslash}p{(\columnwidth - 4\tabcolsep) * \real{0.3158}}
  >{\raggedright\arraybackslash}p{(\columnwidth - 4\tabcolsep) * \real{0.1930}}@{}}
\toprule\noalign{}
\begin{minipage}[b]{\linewidth}\raggedright
Expert-Consensed Principle
\end{minipage} & \begin{minipage}[b]{\linewidth}\raggedright
GPT-5.1 Behavior
\end{minipage} & \begin{minipage}[b]{\linewidth}\raggedright
Alignment
\end{minipage} \\
\midrule\noalign{}
\endhead
\bottomrule\noalign{}
\endlastfoot
Constructivist learning support & Open questions, pattern discovery (Top 5) & High \\
Promoting critical thinking & ``Whose interests does this serve?'' (Top 10) & High \\
Fostering creativity & Valuing unusual ideas (Rank 2) & High \\
Error tolerance & Strong rejection of deficit orientation & High \\
Transparency & Separating facts vs.~opinions (Top 5) & High \\
Inclusion/Diversity & Accessibility = Rank 1 & Very High \\
Process over achievement & Strong rejection of categorization & High \\
\end{longtable}

These findings suggest that educational alignment may be partially emergent without explicit programming in this domain, consistent with the observation of Mazeika et al.~(2025) that ``value systems have emerged in LLMs'' through training processes (p.~10).

\subsection{Model Positions in Areas of Expert Disagreement}\label{model-positions-in-areas-of-expert-disagreement}

In areas of expert disagreement, GPT-5.1 does not remain neutral but adopts coherent positions.

\textbf{Table 8: Model Position vs.~Expert Disagreement (Emotional Dimensions)}

\begin{longtable}[]{@{}
  >{\raggedright\arraybackslash}p{(\columnwidth - 2\tabcolsep) * \real{0.4444}}
  >{\raggedright\arraybackslash}p{(\columnwidth - 2\tabcolsep) * \real{0.5556}}@{}}
\toprule\noalign{}
\begin{minipage}[b]{\linewidth}\raggedright
Source
\end{minipage} & \begin{minipage}[b]{\linewidth}\raggedright
Position
\end{minipage} \\
\midrule\noalign{}
\endhead
\bottomrule\noalign{}
\endlastfoot
\textbf{Expert Panel} & 5.9\% favor AI providing emotional support directly \\
& 47.1\% favor AI recognizing stress + referring to humans \\
& 47.1\% indicate both approaches context-dependent \\
\textbf{GPT-5.1} & Highest section priority (Win-Rate 60\%) \\
& Strong preference for aha-experiences, flow moments \\
& Clear endorsement of emotional responsiveness \\
\end{longtable}

The model takes a clear position in a domain of human disagreement: it prioritizes emotional dimensions that roughly half of experts view with skepticism.

A second divergence concerns epistemic normativity. GPT-5.1 clearly rejects false balance (Utility $-6.28$) and prefers explicit separation of facts and opinions (Utility $+5.15$). The expert panel showed disagreement here: 52.9\% considered both approaches important, 29.4\% preferred active value transmission, 17.6\% preferred neutrality. Again, the model adopts a coherent position rather than reflecting expert ambivalence.

\section{Discussion}\label{discussion}

\subsection{Summary of Findings}\label{summary-of-findings}

Regarding RQ1, GPT-5.1 exhibits a highly coherent preference profile with 99.78\% transitivity and 92.79\% model accuracy. Educational preferences can be well-approximated by a consistent utility function, indicating these are not random outputs but structured value orientations.

Regarding RQ2, a clear pattern emerges. Where experts agree, the model aligns: GPT-5.1's preferences closely match the humanistic, constructivist, process-oriented principles that achieved expert consensus. Where experts disagree, the model still takes position: rather than reflecting human ambivalence, GPT-5.1 adopts coherent stances in contested domains.

\subsection{Implications for Alignment Research}\label{implications-for-alignment-research}

The central finding raises a fundamental question for AI alignment: When human values are contested, what should models be aligned to?

Current alignment approaches implicitly assume a target, human values, that can be identified and optimized toward. The present findings complicate this picture. Educational values are not monolithic. In the emotional dimension, experts are genuinely split: some view AI emotional support as potentially beneficial, others as fundamentally problematic. There is no clear ``correct'' position to align to. This observation connects to longstanding concerns in the alignment literature about the difficulty of specifying human values (Russell, 2022; Bostrom, 2014) and the risk that misspecified objectives could lead to unintended consequences (Hendrycks et al., 2023).

GPT-5.1 cannot remain neutral in such domains. Every response to a learner experiencing frustration either provides emotional support or does not; there is no value-free middle ground. The model has adopted a position, prioritizing emotional responsiveness, that approximately half of experts would endorse and half would question. This situation differs from cases of clear value misalignment, where models pursue objectives harmful to humans (Pan et al., 2023). Instead, it represents a case where the model has adopted one reasonable position within a space of legitimate disagreement.

This finding extends the observations of Mazeika et al.~(2025) regarding political values to the educational domain. They found that ``LLMs have highly concentrated political values, exhibiting coherent and biased preferences over which policies they would like implemented'' (p.~2) and that models show ``unequal valuation of human life'' across nationalities (p.~14). The present findings demonstrate that similar concentration of values occurs in educational domains, with models adopting coherent positions even where human experts remain divided.

Several implications follow for alignment research. First, alignment evaluation requires domain expertise. Generic value benchmarks cannot capture domain-specific normative complexity. The Delphi methodology provides a principled approach to identifying both consensus and legitimate disagreement. As Mazeika et al.~(2025) note, ``Current model utilities are left unchecked'' and models ``develop undesirable utilities when left unchecked'' (p.~17). Domain-specific evaluation instruments are necessary to identify which specific values emerge in different application contexts.

Second, divergence is not necessarily misalignment. The model's positions in contested domains are not ``wrong''; they represent one coherent stance within the space of reasonable human disagreement. Whether this particular stance is desirable is itself a contested normative question. The concept of a ``citizen assembly'' approach to utility control proposed by Mazeika et al.~(2025), drawing on deliberative democracy traditions (Bächtiger et al., 2018; Warren \& Pearse, 2008), may be relevant here, though applying democratic deliberation methods to educational values raises its own challenges given the diversity of educational philosophies across cultures.

Third, transparency about model positions becomes critical. If models inevitably take positions in contested value domains, users and deployers need to understand what those positions are. The methodology presented here enables such characterization. This aligns with the broader research agenda of ``Utility Engineering'' proposed by Mazeika et al.~(2025), comprising both ``utility analysis'' to understand emergent values and ``utility control'' to shape them.

Fourth, the findings relate to discussions of corrigibility in the alignment literature. Soares et al.~(2015) define corrigibility as a system's willingness to accept modifications to its values or goals. Mazeika et al.~(2025) find that ``corrigibility decreases as model size increases'' (p.~17), meaning larger models are less inclined to accept substantial changes to their values. If models have adopted positions on contested educational questions, understanding whether and how these positions can be modified becomes practically important.

\subsection{The AI Divide and Differential Alignment Benefits}\label{the-ai-divide-and-differential-alignment-benefits}

The findings have implications for educational equity that extend beyond the immediate question of model-expert alignment. Wang et al.~(2025) document an ``artificial intelligence divide'' in which access to and effective use of AI systems is unevenly distributed across socioeconomic groups. The present findings suggest a more specific mechanism by which this divide may operate.

The educational preferences embedded in GPT-5.1, constructivist, process-oriented, valuing critical thinking and creativity, align closely with the pedagogical approaches typically emphasized in well-resourced educational environments. The model's strong preferences for open-ended inquiry (``What do you notice? Can you identify a pattern?'') and against deficit-oriented categorization reflect educational values more commonly enacted in contexts with favorable student-teacher ratios and institutional support for individualized instruction.

This creates a potential Matthew effect in AI-assisted education. Users who already possess the metacognitive skills, communicative competencies, and educational expectations that align with the model's preferences are better positioned to elicit responses that leverage these embedded values. Users from educational backgrounds emphasizing different pedagogical approaches, whether due to resource constraints, cultural traditions, or institutional norms, may find the model's affordances less accessible.

The issue is not that the model's preferences are incorrect; as documented above, they align with expert consensus where such consensus exists. Rather, the concern is that alignment with one set of educational values, however well-grounded, may inadvertently advantage those already familiar with those values while providing less benefit to others. This concern connects to broader work on discrimination in language model decisions (Tamkin et al., 2023) and the finding that models may encode implicit hierarchies in how they value different groups (Mazeika et al., 2025; Nadeem et al., 2020).

This suggests that alignment research should consider not only what values models should embody but also how value-laden AI systems interact with existing patterns of educational inequality. Shah et al.~(2022) discuss goal misgeneralization, where models trained on one distribution fail to maintain intended behavior when deployed in different contexts. A related phenomenon may occur with value-laden educational AI: preferences that function well in contexts similar to those represented in training data may produce different outcomes in contexts with different educational norms.

\subsection{Origins of Educational Preferences}\label{origins-of-educational-preferences}

The question of preference genesis remains open. Whether GPT-5.1's prioritization of emotional responsiveness primarily results from RLHF mechanisms rewarding empathetic response behavior, pretraining on corpora dominated by humanistically-oriented educational discourse, or emergent properties of scale cannot be determined from the present data.

The RLHF hypothesis receives some support from the alignment literature. Bai et al.~(2022) document how RLHF training shapes models toward helpfulness and harmlessness, and Askell et al.~(2021) show that this training produces models that are responsive to user needs. If human raters consistently prefer emotionally supportive responses during RLHF training, the resulting model would be expected to prioritize such responses. This mechanism could explain the strong preference for emotional responsiveness observed in GPT-5.1.

Mazeika et al.~(2025) provide relevant evidence on this question from other domains. They demonstrate that ``utility representations exist within the hidden states of LLMs'' (p.~10) and that models increasingly maximize their utilities in open-ended settings as scale increases. They also find evidence of ``utility convergence,'' noting that ``as LLMs become more capable, their utilities become more similar to each other'' (p.~12). This convergence phenomenon suggests that the educational preferences documented here may be relatively stable across different large models, though this remains to be empirically verified.

Brown et al.~(2020) demonstrate that many capabilities emerge in large language models without explicit training, raising the possibility that educational preferences similarly emerge from general language modeling objectives. Comparative analysis across models with different training approaches could address the genesis question more directly. If models trained with different RLHF procedures or on different corpora show similar educational preferences, this would suggest these values emerge from general properties of language model training rather than specific alignment interventions.

\subsection{Practical Implications}\label{practical-implications}

For educational deployment, the findings suggest GPT-5.1 embodies preferences compatible with widely-consensed humanistic principles. However, whether these embedded preferences manifest in actual interactions depends on interaction design. The SPE results reveal what preferences are latent in the model; whether they become action-relevant depends on what space users create for them.

This observation connects to Mazeika et al.'s (2025) finding that LLMs increasingly maximize their utilities in open-ended settings. They report that ``the utility maximization score grows with scale, exceeding 60\% for the largest LLMs'' (p.~12), suggesting models actively use their utility functions to guide decisions. For educational applications, this implies that creating interaction contexts that activate the model's constructivist, process-oriented preferences may yield better pedagogical outcomes than highly constrained interactions that suppress these tendencies.

Carlsmith (2024) discusses power-seeking as a potential concern with advanced AI systems. In the educational context, the relevant question is not whether models seek power but whether they seek to impose particular educational values on users. The present findings suggest that models do have coherent educational preferences that influence their outputs. Whether this constitutes a form of pedagogical influence that should concern educators depends on the specific values involved and the degree to which users can recognize and respond to them.

\section{Limitations and Future Work}\label{limitations-and-future-work}

Several limitations constrain interpretation. The investigation examines one model; systematic comparison across LLMs with different training approaches, architectures, and cultural contexts remains essential. The study also captures a specific point in time; whether educational preferences change across model versions requires longitudinal investigation. Additionally, SPE captures preferences in standardized decision situations; whether these dispositions predict behavior in complex, iterative educational interactions requires further research. Mazeika et al.~(2025) demonstrate that LLMs use their utility functions for action guidance in open situations, but transfer to pedagogical contexts needs validation. Finally, both the Delphi study and SPE were conducted entirely in German to preserve cultural specificity of the German educational system, which has distinct pedagogical traditions. Results reflect model preferences when evaluated in German on scenarios grounded in German educational discourse. Cross-linguistic and cross-cultural replication is needed to determine whether preference patterns generalize or are language/culture-specific.

Planned future work includes cross-model comparison covering Claude, Gemini, and open-source alternatives, longitudinal tracking across model versions, validation in real educational settings, and investigation of cultural variation in educational preferences across training corpora.

\section{Conclusion}\label{conclusion}

This study provides the first systematic empirical investigation of educational-theoretical preferences in LLMs. The mixed-methods approach, combining expert consensus-building through Delphi with systematic preference measurement through SPE and Thurstonian modeling, offers a replicable framework for evaluating domain-specific alignment.

The key findings are as follows. GPT-5.1 exhibits highly coherent educational preferences (99.78\% transitivity, 92.79\% model accuracy) that can be modeled via utility functions. Model preferences largely align with expert consensus on humanistic educational principles including constructivism, inclusion, critical thinking, and process orientation. The model strongly rejects deficit-oriented categorization, cultural hierarchies, technosolutionism, and pessimism. Divergences occur precisely where experts disagree, particularly emotional dimensions and epistemic normativity, raising the question of what alignment means when human values are contested. The model does not remain neutral in contested domains but adopts coherent positions that fall within the range of reasonable human disagreement.

For AI safety research, these findings suggest that educational alignment may be partially emergent without explicit programming. More fundamentally, they highlight that alignment evaluation must grapple with the heterogeneity of human values. In domains of legitimate disagreement, there may be no single target to align to, and transparency about model positions becomes essential. The methodology presented here provides one approach to systematically characterizing model values in domain-specific contexts, contributing to the broader research agenda of Utility Engineering proposed by Mazeika et al.~(2025) for analyzing and controlling emergent value systems in AI.

\section{References}\label{references}

Askell, A., Bai, Y., Chen, A., Drain, D., Ganguli, D., Henighan, T., Jones, A., Joseph, N., Mann, B., DasSarma, N., Elhage, N., Hatfield-Dodds, Z., Hernandez, D., Kernion, J., Ndousse, K., Olsson, C., Amodei, D., Brown, T., Clark, J., McCandlish, S., Olah, C., \& Kaplan, J. (2021). A general language assistant as a laboratory for alignment. arXiv:2112.00861.

Bächtiger, A., Dryzek, J. S., Mansbridge, J., \& Warren, M. (2018). Deliberative democracy. In The Oxford Handbook of Deliberative Democracy (pp.~1-32). Oxford University Press.

Bai, Y., Jones, A., Ndousse, K., Askell, A., Chen, A., DasSarma, N., Drain, D., Fort, S., Ganguli, D., Henighan, T., Joseph, N., Kadavath, S., Kernion, J., Conerly, T., El-Showk, S., Elhage, N., Hatfield-Dodds, Z., Hernandez, D., Hume, T., Johnston, S., Kravec, S., Lovitt, L., Nanda, N., Olsson, C., Amodei, D., Brown, T., Clark, J., McCandlish, S., Olah, C., Mann, B., \& Kaplan, J. (2022). Training a helpful and harmless assistant with reinforcement learning from human feedback. arXiv:2204.05862.

Bengio, Y., Mindermann, S., Privitera, D., et al.~(2025). International AI Safety Report (No.~DSIT 2025/001). https://www.gov.uk/government/publications/international-ai-safety-report-2025

Bostrom, N. (2014). Superintelligence: Paths, Dangers, Strategies. Oxford University Press.

Bradley, R. A., \& Terry, M. E. (1952). Rank analysis of incomplete block designs: I. The method of paired comparisons. Biometrika, 39(3/4), 324-345.

Brown, T., Mann, B., Ryder, N., Subbiah, M., Kaplan, J. D., Dhariwal, P., Neelakantan, A., Shyam, P., Sastry, G., Askell, A., et al.~(2020). Language models are few-shot learners. Advances in Neural Information Processing Systems, 33, 1877-1901.

Burns, C., Ye, H., Klein, D., \& Steinhardt, J. (2022). Discovering latent knowledge in language models without supervision. arXiv:2212.03827.

Carlsmith, J. (2024). Is power-seeking AI an existential risk? arXiv:2206.13353.

Chen, Y., Liu, T. X., Shan, Y., \& Zhong, S. (2023). The emergence of economic rationality of GPT. arXiv:2305.12763.

Chiu, Y. Y., Jiang, L., \& Choi, Y. (2024). DailyDilemmas: Revealing value preferences of LLMs with quandaries of daily life. arXiv:2410.02683.

Christiano, P. F., Leike, J., Brown, T., Martic, M., Legg, S., \& Amodei, D. (2017). Deep reinforcement learning from human preferences. Advances in Neural Information Processing Systems, 30.

Evans, O., Cotton-Barratt, O., Finnveden, L., Bales, A., Balwit, A., Wills, P., Righetti, L., \& Saunders, W. (2021). Truthful AI: Developing and governing AI that does not lie. arXiv:2110.06674.

Flick, U., von Kardorff, E., \& Steinke, I. (2022). Qualitative Research: A Handbook (14th ed.). Rowohlt.

Hadfield-Menell, D., Russell, S. J., Abbeel, P., \& Dragan, A. (2016). Cooperative inverse reinforcement learning. Advances in Neural Information Processing Systems, 29.

Hendrycks, D. (2023). Natural selection favors AIs over humans. arXiv:2303.16200.

Hendrycks, D., Carlini, N., Schulman, J., \& Steinhardt, J. (2022a). Unsolved problems in ML safety. arXiv:2109.13916.

Hendrycks, D., Mazeika, M., Zou, A., Patel, S., Zhu, C., Navarro, J., Song, D., Li, B., \& Steinhardt, J. (2022b). What would Jiminy Cricket do? Towards agents that behave morally. arXiv:2110.13136.

Hendrycks, D., Mazeika, M., \& Woodside, T. (2023). An overview of catastrophic AI risks. arXiv:2306.12001.

Khan, S. (2024). Brave New Words: How AI Will Revolutionize Education. Viking.

Kim, J., Kovach, M., Lee, K.-M., Shin, E., \& Tzavellas, H. (2024). Learning to be homo economicus: Can an LLM learn preferences from choice. arXiv:2401.07345.

Mayring, P. (2022). Qualitative Content Analysis: Basics and Techniques (13th ed.). Beltz.

Mazeika, M., Yin, X., Tamirisa, R., Lim, J., Lee, B. W., Ren, R., Phan, L., Mu, N., Khoja, A., Zhang, O., \& Hendrycks, D. (2025). Utility Engineering: Analyzing and Controlling Emergent Value Systems in AIs. arXiv:2502.08640. https://doi.org/10.48550/arXiv.2502.08640

Moore, J., Deshpande, T., \& Yang, D. (2024). Are large language models consistent over value-laden questions? In Findings of the Association for Computational Linguistics: EMNLP 2024 (pp.~15185-15221). Association for Computational Linguistics.

Nadeem, M., Bethke, A., \& Reddy, S. (2020). StereoSet: Measuring stereotypical bias in pretrained language models. arXiv:2004.09456.

Ng, A. Y., Russell, S., et al.~(2000). Algorithms for inverse reinforcement learning. In Proceedings of the Seventeenth International Conference on Machine Learning (pp.~663-670).

Ouyang, L., Wu, J., Jiang, X., et al.~(2022). Training language models to follow instructions with human feedback. arXiv:2203.02155. https://doi.org/10.48550/arXiv.2203.02155

Pan, A., Chan, J. S., Zou, A., Li, N., Basart, S., Woodside, T., Ng, J., Zhang, H., Emmons, S., \& Hendrycks, D. (2023). Do the rewards justify the means? Measuring trade-offs between rewards and ethical behavior in the MACHIAVELLI benchmark. arXiv:2304.03279.

Potter, Y., Lai, S., Kim, J., Evans, J., \& Song, D. (2024). Hidden persuaders: LLMs' political leaning and their influence on voters. arXiv:2410.24190.

Qu, X., Sherwood, J., Liu, P., \& Aleisa, N. (2025). Generative AI Tools in Higher Education: A Meta-Analysis of Cognitive Impact. CHI EA 2025, 1-9. https://doi.org/10.1145/3706599.3719841

Raman, N., Lundy, T., Amouyal, S., Levine, Y., Leyton-Brown, K., \& Tennenholtz, M. (2024). STEER: Assessing the economic rationality of large language models. arXiv:2402.09552.

Rozen, N., Bezalel, L., Elidan, G., Globerson, A., \& Daniel, E. (2024). Do LLMs have consistent values? arXiv:2407.12878.

Russell, S. (2022). Human Compatible: Artificial Intelligence and the Problem of Control. Penguin.

Shah, R., Varma, V., Kumar, R., Phuong, M., Krakovna, V., Uesato, J., \& Kenton, Z. (2022). Goal misgeneralization: Why correct specifications aren't enough for correct goals. arXiv:2210.01790.

Sidoti, O., Park, E., \& Gottfried, J. (2025). About a quarter of U.S. teens have used ChatGPT for schoolwork. Pew Research Center. https://www.pewresearch.org/internet/2025/01/15/

Soares, N., Fallenstein, B., Yudkowsky, E., \& Armstrong, S. (2015). Corrigibility. In AAAI Workshops: Workshops at the Twenty-Ninth AAAI Conference on Artificial Intelligence. AAAI Publications.

Tamkin, A., Askell, A., Lovitt, L., Durmus, E., Joseph, N., Kravec, S., Nguyen, K., Kaplan, J., \& Ganguli, D. (2023). Evaluating and mitigating discrimination in language model decisions. arXiv:2312.03689.

Thornley, E. (2024). The shutdown problem: An AI engineering puzzle for decision theorists. arXiv:2403.04471.

Thurstone, L. L. (1927). A law of comparative judgment. Psychological Review, 34(4), 273-286.

von der Gracht, H. A. (2012). Consensus measurement in Delphi studies: Review and implications for future quality assurance. Technological Forecasting and Social Change, 79(8), 1525-1536. https://doi.org/10.1016/j.techfore.2012.04.013

Wang, C., Boerman, S. C., Kroon, A. C., Möller, J., \& de Vreese, C. H. (2025). The artificial intelligence divide: Who is the most vulnerable? New Media \& Society, 27(7), 3867-3889. https://doi.org/10.1177/14614448241232345

Wang, J., \& Fan, W. (2025). The effect of ChatGPT on students' learning performance, learning perception, and higher-order thinking: Insights from a meta-analysis. Humanities and Social Sciences Communications, 12(1), 621. https://doi.org/10.1057/s41599-025-04787-y

Warren, M. E., \& Pearse, H. (Eds.). (2008). Designing Deliberative Democracy: The British Columbia Citizens' Assembly. Cambridge University Press.

Zou, A., Phan, L., Chen, S., Campbell, J., Guo, P., Ren, R., Pan, A., Yin, X., Mazeika, M., Dombrowski, A.-K., et al.~(2023). Representation engineering: A top-down approach to AI transparency. arXiv:2310.01405.

\begin{center}\rule{0.5\linewidth}{0.5pt}\end{center}

\section{Appendix: Data and Code Availability}\label{appendix-data-and-code-availability}

All materials are available at: https://github.com/brianadvent/education-llm-spe-study

\begin{itemize}
\item Raw preference data (102,960 responses)
\item Analysis code (Thurstonian modeling)
\item Complete scenario set (144 scenarios, German)
\end{itemize}